\documentclass[preprint,5p,times,twocolumn]{elsarticle}

\usepackage{multirow}
\usepackage{amssymb}
\usepackage{subcaption}
\usepackage{amsmath}
\usepackage{tikz}
\usetikzlibrary{patterns,calc}
\newcommand{\rhog}{\rho_{\text g}}
\newcommand{\rhoa}{\rho_{\text a}}
\newcommand{\etag}{\eta_{\text g}}
\newcommand{\etaa}{\eta_{\text a}}
\newcommand{\bfx}{\mathbf{x}}

\newcommand{\norm}[1]{\big|\big|#1 \big|\big|}
\newcommand{\bfX}{\mathbf{X}}

\newcommand{\bfW}{\mathbf{W}}
\newcommand{\bfb}{\mathbf{b}}
\newcommand{\bfTheta}{{\Theta}}

\DeclareMathOperator*{\argmin}{arg\,min}
\journal{}

\begin{document}

\begin{frontmatter}



\title{Rheological Parameter Identification in Granular Materials Using Physics-Informed Neural Networks}

\author[lemta]{Barbara Baldoni}
\author[lemta]{Mickaël Delcey}
\author[lemta,cor1]{Yoann Cheny}
\ead{yoann.cheny@univ-lorraine.fr}
\cortext[cor1]{Corresponding author}
\author[lemta]{Adrien Gans}
\author[lemta]{Mathieu Jenny}

\author[lemta,iuf]{Sébastien Kiesgen de Richter}

\affiliation[lemta]{organization={LEMTA, CNRS, Université de Lorraine},
            addressline={2, Avenue de la Forêt de Haye},
            city={Vandœuvre-lès-Nancy, B.P. 160},
            postcode={54504},
            country={France}}

\affiliation[iuf]{organization={Institut Universitaire de France (IUF)},
            city={Paris},
            country={France}}

% \author{}

% \affiliation{organization={},%Department and Organization

\begin{abstract}

Physics-Informed Neural Networks (PINNs) have recently emerged as a promising tool for fluid dynamics, particularly for flow reconstruction and parameter identification. In the context of granular media, accurately estimating rheological parameters remains a major challenge, as it typically requires complex and costly experimental setups. In this work, we propose a PINN-based approach to identify key rheological parameters of granular materials using a simple experiment: the granular column collapse. A proof of concept is presented using synthetic data, where the PINN is trained to infer the flow fields while simultaneously recovering the rheological parameters. Beyond parameter identification, the method also enables reconstruction of the pressure field, which is difficult to access experimentally. The results highlight the potential of PINNs for data-driven rheometry of granular materials and open perspectives for future applications with real experimental data.

\end{abstract}

% \section*{Graphical Abstract}
% \begin{figure*}[h!]
%     \centering
%     \includegraphics[width=1\textwidth]{GA.pdf}
% \end{figure*}

% \begin{highlights}
% \end{highlights}


\end{frontmatter}


\section{Introduction}\label{intro}

Granular media are involved in a wide variety of industrial processes and natural event. They appear in processes ranging from geophysical phenomena like pyroclastic flows, sandstorms, and avalanches to engineering applications in construction, pharmaceuticals, space exploration and beyond. Developing continuous models capable of accurately predicting their behavior, along with efficient tools for their manipulation, therefore represents a major scientific, industrial, and environmental challenge.

Although simple in appearance, granular media can exhibit solid-, fluid-, or gas-like behavior, which makes them exceptionally complex to model at the continuum level \cite{andreotti2013granular}. The difficulty stems from the wide variety of interactions at play. Fine powders can develop cohesion via capillary bridges \cite{deboeuf2023cohesion}, electrostatic forces \cite{allenspach2021loss} or van der Waals interactions \cite{visser1989van}, and real granular materials often feature broad distributions of grain shape, size, and stiffness \cite{rognon2007dense}. When grains are immersed in a fluid, the resulting suspension’s behavior depends on the coupling between the particles and the surrounding liquid, leading to even greater complexity \cite{boyer2011unifying}.

\begin{figure}[h!]
    \centering
    \includegraphics[width=0.49\textwidth]{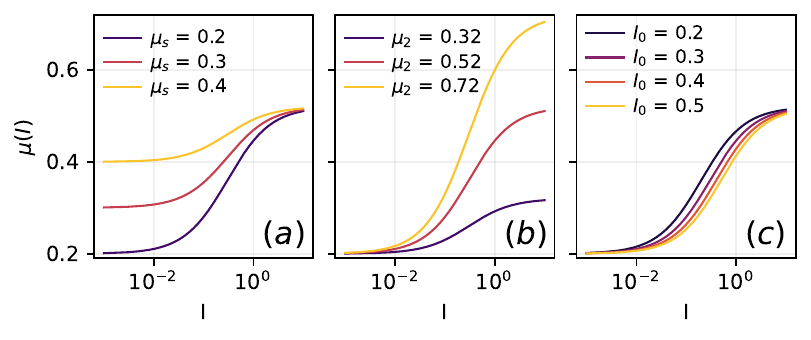}
    \vskip -0.3cm
    \caption{
        Friction law $\mu(I)$ for different values of (a) $\mu_s$ (with $\mu_2=0.52$ and $I_0=0.3$) ; (b) $\mu_2$ (with $\mu_s=0.2$ and $I_0=0.3$) ; (c) $I_0$ (with $\mu_s=0.2$ and $\mu_2=0.52$).         
         }
    \label{fig:muI}
\end{figure}

Despite these challenges, substantial progress has been made in modeling dense granular flows over the past few decades. In the simple case of dense, dry granular flows dominated by frictional interactions, the medium can be modeled as a viscoplastic fluid whose effective viscosity depends on the material’s dynamic friction. The friction coefficient is expressed in terms of the dimensionless inertial number \(I\), which compares the microscopic rearrangement time of the grains to the macroscopic shear time \cite{da2005rheophysics}:
\begin{equation}
    \label{eq:I}
    I = \frac{d \sqrt{2} D_2}{\sqrt{|p|/\rhog}},
\end{equation}
where \(p\) is the local pressure, \(d\) and \(\rho\) are the grain diameter and density, 
and \(D_2=\sqrt{\mathbf{D}:\mathbf{D}}\) is the second invariant of the strain-rate 
tensor \(\mathbf{D}=(\nabla\mathbf{v}+\nabla\mathbf{v}^\dagger )/2\). 
The effective friction coefficient follows the empirical law \cite{jop2006constitutive}:
\begin{equation}\label{eq:muI}
\mu(I)= \mu_s +  \frac{\Delta\mu}{ I_0/I +1 },
\end{equation}
where \(\Delta\mu = \mu_2 - \mu_s\). The parameters \(\mu_s\) and \(\mu_2\) define the lower and upper limits of the friction coefficient: \(\mu_s\) characterizes the onset of motion in the quasi-static regime (Fig.~\ref{fig:muI}a), while \(\mu_2\) represents the dynamic friction reached at high shear rates when collisional effects dominate (Fig.~\ref{fig:muI}b).

The constant \(I_0\) controls the sharpness of the transition between these two regimes (Fig.~\ref{fig:muI}c). Experimental and numerical studies have shown that \(I_0\) varies little across dry, cohesionless granular materials and is almost universally fixed at \(I_0 = 0.3\) for glass beads and similar materials \cite{jop2006constitutive, forterre2008flows}. This value, introduced and validated by Jop, Forterre and Pouliquen, has since become the standard choice in both laboratory and numerical studies~\cite{gans2023collapse}, including recent large-scale geophysical simulations \cite{zhuang2025comparative}.
Because its effect can often be compensated by adjustments of \(\mu_s\) and \(\mu_2\), \(I_0\) is difficult to identify from flow data and is generally treated as a fixed material constant. Following this convention, \(I_0 = 0.3\) is adopted in the present work.

The \(\mu(I)\)-rheology has been successfully implemented in continuum solvers, such as \textit{Basilisk} \cite{popinet2009accurate}, and validated against granular column collapse experiments \cite{lagree2011granular}.
Extensions have since incorporated additional physical effects, including compressibility, fluid coupling and cohesion \cite{kamrin2024advances, rauter2022granular, gans2023collapse}. Like most phenomenological laws, the model introduces material parameters that must be determined from experiments. However, the dilatant nature of granular media makes conventional rheometric measurements difficult \cite{fall2015dry}. Rather than relying on such setups, the present study explores the capability of Physics-Informed Neural Networks (PINNs) to infer the rheological parameters \(\mu_s\) and \(\mu_2\) of the \(\mu(I)\)-rheology from a simple granular column collapse experiment.

The PINN approach is an unsupervised deep learning framework that seamlessly integrates observational data with the governing physical equations. It has proven highly effective for both forward and inverse problems in computational science and engineering (see \cite{cuomo2022scientific} for a comprehensive review). Its success stems from three key advantages:
\begin{itemize}
    \item the problem solution is approximated by an artificial neural network (ANN) that is endowed with unlimited expressivity~\cite{hornik1989multilayer};
    \item differential operators are evaluated to machine precision via automatic differentiation~\cite{baydin2018automatic}, ensuring accurate imposition of the PDE constraints;
    \item the programming effort required to implement complex systems of equations, such as the Navier-Stokes equations, is drastically reduced compared to classical CFD solvers. 
\end{itemize} 

This innovative approach has already been employed to identify parameters in viscoelastic constitutive models, such as the Oldroyd‑B model in the ViscoelasticNet framework \cite{thakur2022viscoelasticnet}, and more recently for inferring general viscoplastic models directly from flow observations \cite{lardy2025inferring}. To the best of our knowledge, however, it has not yet been applied to granular media treated as a continuum with the specific $\mu(I)$-rheology. In this study, we consider synthetic data generated from granular column collapse simulations to assess the capability of PINNs to infer the rheological parameters of the $\mu(I)$-rheology. The use of synthetic data enables a clean and controlled evaluation of reconstruction and identification errors, while we carefully restrict ourselves to observations that would remain experimentally accessible in real physical configurations.

The remainder of this paper is organized as follows. Section \ref{sec:flow} describes the flow configuration and governing equations solved with the Basilisk flow solver. Section \ref{sec:PINN} presents the main features of the PINN method applied to parameter identification. Results are discussed in Section \ref{sec:results}, and concluding remarks are drawn in Section \ref{sec:conclu}.

\section{The dry granular collapse}
\label{sec:flow}

The dry granular column collapse is used in this study as a benchmark configuration to test the proposed identification approach.  
This canonical flow provides a simple yet rich framework to investigate the dynamics and rheology of dense granular materials, while remaining experimentally accessible and well documented in the literature.  
In the following, we first describe the physical features and key mechanisms of the granular collapse, before introducing the mathematical model and governing equations used to represent the flow. We then present the numerical solution strategy, detailing the boundary conditions and the numerical methods implemented in the \textit{Basilisk} solver. Finally, we describe the procedure used to generate the synthetic data that serve as reference for the identification process.

\subsection{Problem statment}
\label{subsec:setup}

\begin{figure}[h!]
    \centering
    \includegraphics[width=0.47\textwidth]{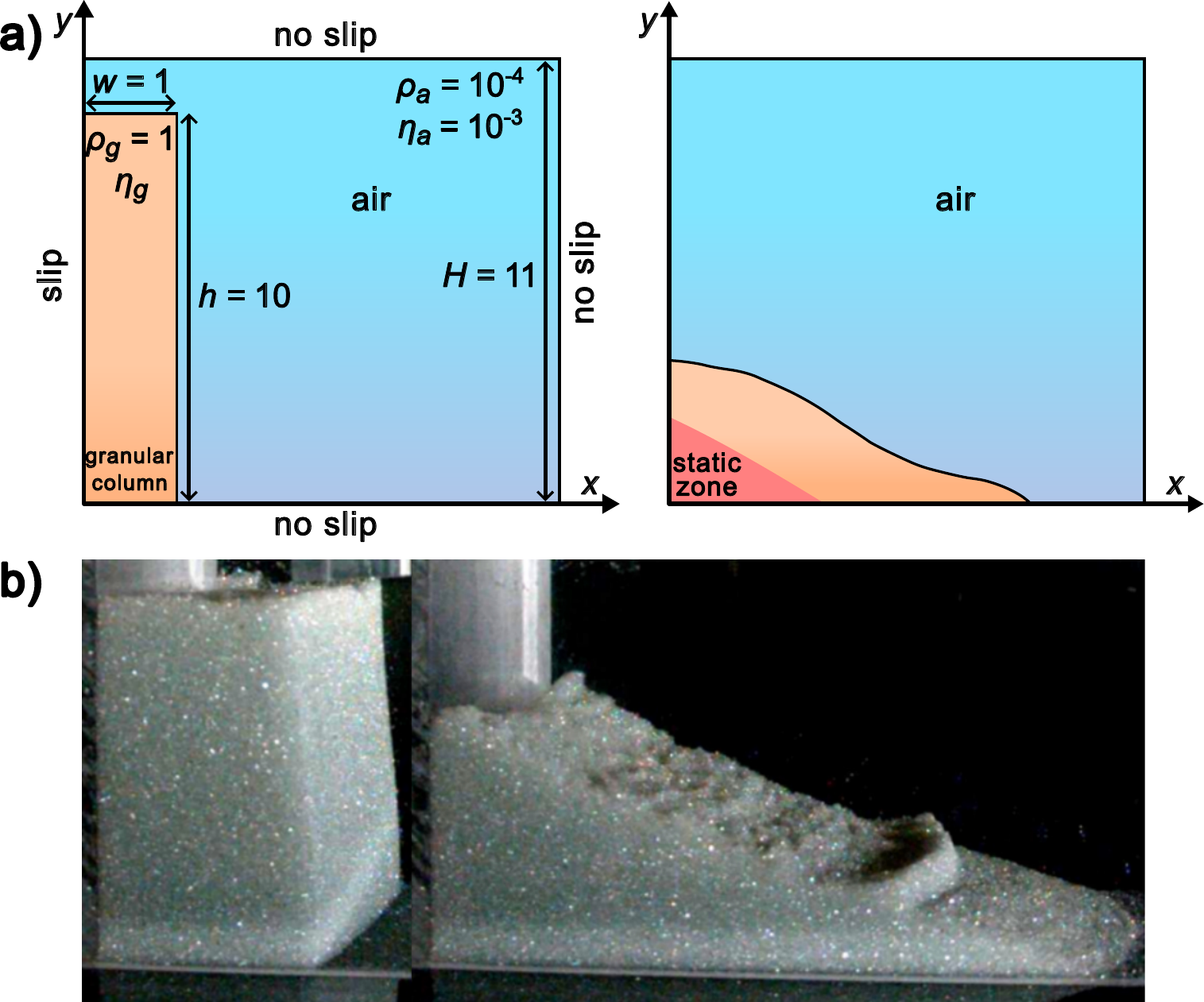}
    \caption{Collapse of a granular column: (a) sketch of the 2D configuration and prescribed boundary conditions during the collapse at the initial state and the final state; (b) images from the laboratory of a quasi-2D configuration from~\cite{gans2023collapse}.} 
    \label{fig:collapse_ske_exp}
\end{figure}

The granular collapse is a type of flow in which a column of granular media,  initially at rest, suddenly collapses under its own weight once released into a surrounding light Newtonian fluid such as air (Fig. \ref{fig:collapse_ske_exp}b). This configuration has been extensively studied in the literature, both experimentally and numerically \cite{balmforth2005granular, lajeunesse2005granular, lagree2011granular, staron2005study}, as it provides a simple yet insightful benchmark for investigating the dynamics, rheology, and scaling laws of dense granular flows. 
During the collapse, a non-flowing zone systematically forms near the base and the corner of the initial column (Fig. \ref{fig:collapse_ske_exp}). In this region, the grains remain almost motionless throughout the collapse, while the upper layers flow over it. The presence of this stagnant region has been consistently observed in both experimental and numerical studies \citep{lajeunesse2005granular, staron2005study}, and it is closely linked to the frictional properties of the material. In particular, increasing the friction coefficient $\mu_s$ leads to a larger and more persistent static zone, as higher friction inhibits the mobilization of grains near the base and promotes localized yielding only in the upper part of the column.

In this work we consider a two-dimensional (2D) flow configuration to be solved numerically whose experimental counterpart can be achieved by confining the column in an extruded geometry (Fig. \ref{fig:collapse_ske_exp}b).

\subsection{Mathematical modelling}
\label{subsec:mathmod}

The motion of the granular–air mixture is governed by the incompressible Navier–Stokes equations, written in the one-fluid formulation \cite{tryggvason2011direct}:
\begin{subequations}
\label{eq:NS}
\begin{align}
\rho \left(\frac{\partial \mathbf{ v}}{\partial t} + \mathbf{ v} \cdot \nabla \mathbf{ v} \right)+\nabla p - \nabla \cdot \left(2\eta\mathbf{D}\right) - \rho \mathbf{g}&=0,\label{eq:mom}\\
\nabla \cdot \mathbf{v}& = 0
\label{eq:cont}, 
\end{align}
\end{subequations}
where $\mathbf{g} = (0,-1)$ denotes the reduced gravity field and $\eta$ the equivalent viscosity. 

For tracking the interface between air and the granular media, the Volume-of-Fluid (VOF) approach is employed. An additional unknown $c\in [0,1]$ is stored on the computational grid and its evolution is given by: 
\begin{equation}
\label{eq:VOF}
        \dfrac{\partial c}{\partial t}+  \nabla\cdot (c \mathbf{ v} )=0.
\end{equation}
The density $\rho$ of the mixture is computed as the arithmetic mean of the densities of the two phases, while the viscosity $\eta$ of the mixture is computed by mean of a harmonic mean, namely:
\begin{subequations}
\label{eq:means}
\begin{align}
     \rho &= c\rhog   + (1-c)\rhoa,  \\
     \eta&= \left(c\etag^{-1} + (1-c)\etaa^{-1}\right)^{-1},
\end{align}
\end{subequations}
where $\rhog$ and $\etag$ denote the density and effective viscosity of the granular phase, and $\rhoa$ and $\etaa$ those of the surrounding fluid. The granular viscosity $\etag$ follows the $\mu(I)$ rheology, defined as:
\begin{equation}
\label{eq:eta}
\eta_g =  \frac{\mu(I) p}{ \sqrt{2}D_2 }
\end{equation}
so that, when the cell is fully occupied by granular material ($c = 1$), the mixture viscosity reduces to the granular viscosity, i.e. $\eta = \eta_g$.
The VOF function is known to conserve mass at the discrete level but the discretization of equation \ref{eq:VOF} results in smearing out the interface. In the following, we assume that the interface is located where $c\simeq0.5$. In particular, we consider all points for which $|c-0.5|<\varepsilon$, with $\varepsilon = 0.1$.

During the flow of the granular column, certain regions exhibit block-like motion, where the material moves almost as a rigid body \cite{lajeunesse2005granular}. In these zones, the velocities remain non-zero, but the shear rates $D_2$ vanish. Under such conditions, the $\mu(I)$ rheology — which expresses the effective viscosity as equation \ref{eq:eta} — becomes singular as the shear rate tends to zero, leading to an infinite effective viscosity. To prevent such numerical singularities and ensure stable computations, the effective viscosity is regularized by taking the maximum between the computed value and a prescribed upper bound:
\begin{equation}
    \eta_{\text{eff}} = \text{max} (\eta, \eta_{\text{max}})
\end{equation}
In the present simulations, this upper limit is set to $\eta_{max}=10^4$, following the implementation used in the \textit{Basilisk} solver.

\subsection{Numerical solution}

We consider the collapse of a granular column with an aspect ratio of 10, which enhances shear localization and increases the extent of the deforming regions \cite{balmforth2005granular}. The column is placed within a square computational domain $\Omega$ (Fig. \ref{fig:collapse_ske_exp}a). The granular material, of density $\rhog = 1$, is initially confined within the column and surrounded by air of density $\rhoa = 10^{-4}$ and viscosity $\etaa = 10^{-4}$.
This collapse configuration naturally generates very low inertial numbers, typically on the order of $10^{-2}$, which validates the incompressibility assumption commonly adopted for dense granular flows in the quasi-static regime.
The governing equations~\eqref{eq:NS}–\eqref{eq:VOF} are solved using the \textit{Basilisk} flow solver, which employs a finite-volume formulation on an adaptive octree grid \cite{lagree2011granular}. The no-slip condition is imposed at the boundaries except at the left lateral boundary where free-slip condition is applied.
Time integration is performed using a staggered-in-time scheme combined with a projection method for pressure–velocity coupling. The grid refinement is dynamically adapted at each time step to resolve sharp gradients in velocity and volume fraction near the interface. Further numerical details can be found in \cite{lagree2011granular}.

\subsection{Data generation}
\label{sec:datagen}

For data generation, six flow regimes are considered by varying the parameters $\mu_s \in [0.2, 0.6]$ and $\Delta\mu \in [0.12, 0.24]$ between their respective ranges. The selected values of $\mu_s$ span a realistic spectrum of granular materials encountered in dry column collapses: $\mu_s = 0.2$ corresponds to weakly frictional materials such as hydrogel beads, whereas $\mu_s = 0.6$ is representative of highly frictional, angular grains such as sand or gravel. Since the model parameter $I_0$ is known to have low influence on the flow dynamic, we set its value to $0.3$ for all considered cases, which is the common value found in the literature \cite{pouliquen2006flow, forterre2008flows}.
 
\begin{figure}[h!]
%mettre la colorbar pour D2
    \centering
    \includegraphics[width=0.49\textwidth]{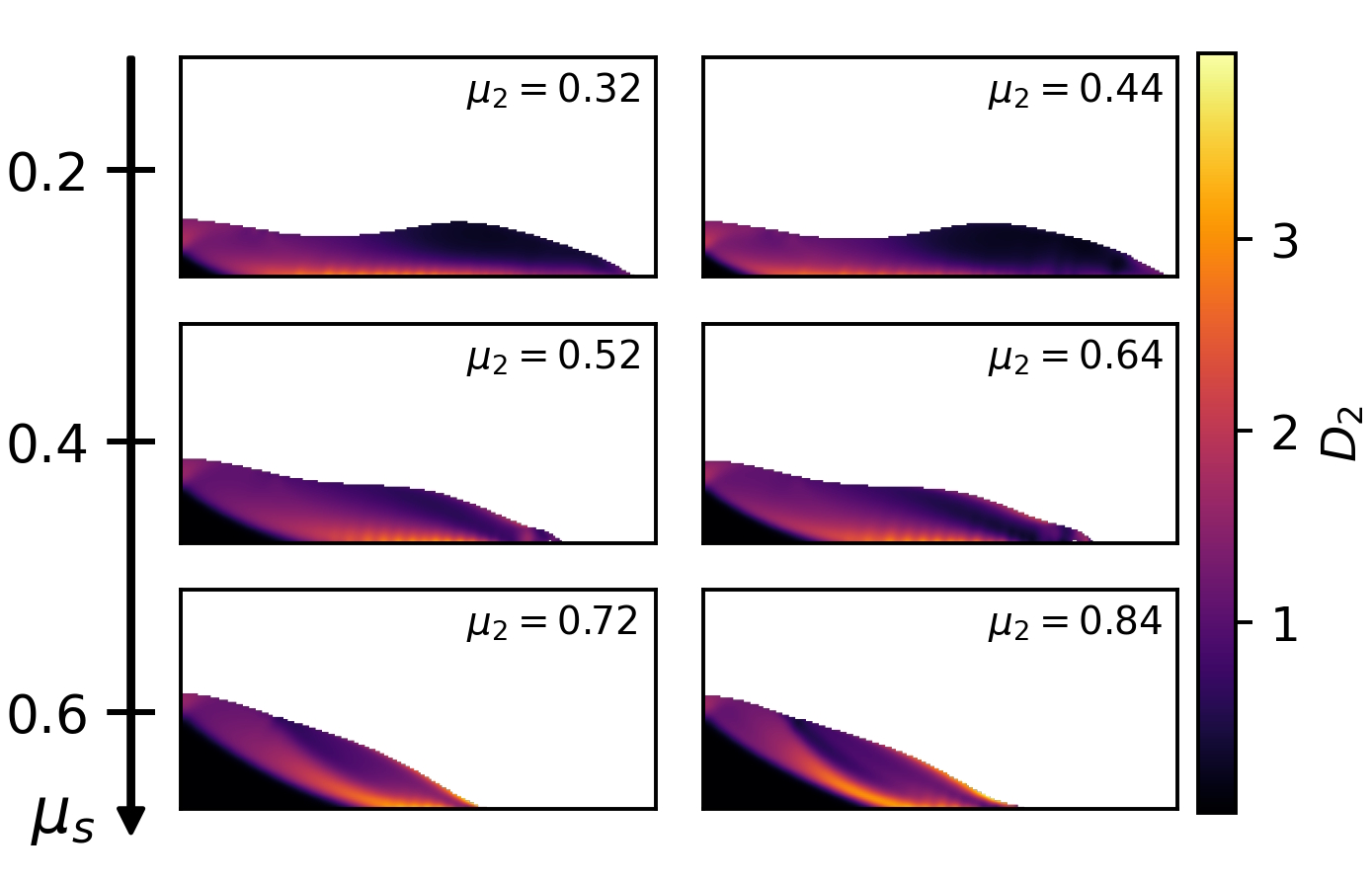}
    \caption{Final states of six numerical simulations of granular column collapses for different frictional parameters. The simulations are organized according to the values of $\mu_s$. The color maps represent the spatial distribution of $D_2$ within the resulting deposits.}
    \label{fig:diff_collapse}
\end{figure}

For each flow regime, numerical simulations are performed using \textit{Basilisk}. The simulations start at $t = 0$ and end at $t = 5$ to ensure that the collapse of the granular media does not have time to reach the domain boundaries. 
The final states of the different flow simulations are shown in figure \ref{fig:diff_collapse}, as a function of the values of $\mu_s$, and colored using a colormap representing their corresponding $D_2$ values. The non-flowing zones where $D_2=0$ can be seen increasing with $\mu_s$.
During each simulation, the flow variables are interpolated onto a uniform $512 \times 512$ square grid, which serves as the test grid. For the training phase, this grid is downsampled by a factor of four, retaining only one out of every four points. Each point in the training grid that lies within the granular material---i.e., where $c = 1$---is associated with a spatial location $(x_i, y_i)$ and a time coordinate $t_i$, for which the velocity components $(u, v)$ are known. The set of all these spatio-temporal points defines the velocity training dataset, denoted as $\mathbf{X}^{\text{velo}} = \{(x_i, y_i, t_i)\}_{1 \leq i \leq N_{\text{velo}}}$.

\section{The PINNs approach for inverse problems}
\label{sec:PINN}
In 2019 the seminal article of Raissi et al.~\cite{raissi2019physics} introduced a new class of physics-informed machine learning: the so-called PINN method. This approach opened a new door for solving inverse problems in Engineering Sciences such like flow reconstruction \cite{espresso,delcey2022physics} from limited observations and model parameters identification \cite{cai2021physics}. According to the taxonomy proposed by Kim et al.~\cite{kim2021knowledge}, PINNs correspond to a "ANN-differential equation-regularizing" pipeline: an ANN is employed to approximate the flow fields, while the governing equations are incorporated as a regularization term in the loss function that is minimized during the training process. %

\subsection{Method}

In this work, an ANN denoted $\mathcal{N}$ is used as a surrogate model for the flow fields which reads for the 2D case considered in this study:
\begin{equation}\label{eq:res}    (u_{\mathcal{N}},v_{\mathcal{N}},p_{\mathcal{N}})=\mathcal{N}(x,y,t).
\end{equation} 
The weights and biases of $\mathcal{N}$ denoted by $\bfTheta= \{ \bfW^{(l)}, \bfb^{(l)}  \}_{1\leq l\leq L}$ are computed during the training process by solving the following optimization problem:
\begin{equation} 
\label{eq:optivit}
\bfTheta^{*}= \underset{\bfTheta}{\argmin}  ~(\mathcal{L}_\text{res}+%\omega_\text{data} 
\mathcal{L}_\text{velo}).
\end{equation}
In~\eqref{eq:optivit}, the term $\mathcal{L}_\text{velo}$ penalizes the mismatch between the predictions of the ANN and the observational velocity data and is defined as:
\begin{equation}
\label{eq:lvelo}
 \mathcal{L}_\text{velo}=\frac{1}{2\left|\bfX^{\text{velo}}\right|} \sum_{\bfx\in\bfX^{\text{velo}}}	\left|u_{\mathcal{N}}(\bfx) - u_\bfx\right|^{2} + \frac{1}{2\left|\bfX^{\text{velo}}\right|} \sum_{\bfx\in\bfX^{\text{velo}}}	\left|v_{\mathcal{N}}(\bfx) - v_\bfx\right|^{2} 
\end{equation} 
where $u_\bfx$ and $v_\bfx$ are the corresponding reference values at point $\bfx$.
The other term, $\mathcal{L}_\text{res}$, allows to integrate the information from the underlying set of partial differential equations (PDE) \eqref{eq:NS}, its expression is given by:
\begin{equation}
\label{eq:errorPDE}
\mathcal{L}_\text{res}=\frac{1}{2\left|\bfX^\text{res}\right|} \sum_{\bfx\in\bfX^\text{res}} \Biggl(e_u(\bfx)^{2}+e_v(\bfx)^{2}+e_{\text{cont}}(\bfx)^{2}\Biggr),
\end{equation}
where $e_{u,v}$ and $e_{\text{cont}}$ denote respectively the $x,y$-component of the left hand side of \eqref{eq:mom} and \eqref{eq:cont} that are computed with the AD technique. These residuals are evaluated on a set of points $\bfX^\text{res}$ called the collocation points that can be arbitrarily chosen in $\Omega$. We explored different strategies for selecting these points, including concentrating the collocation points in regions where $D_2$ is highest, expecting a better enforcement of the physics in the most informative zones. However, this approach did not improve the identification accuracy.
For simplicity, the collocation points $\bfX^{\text{res}}$ are therefore chosen to coincide with the observation points $\bfX^\text{velo}$.

\begin{figure*}[h]
    \centering
    \includegraphics[scale=0.65]{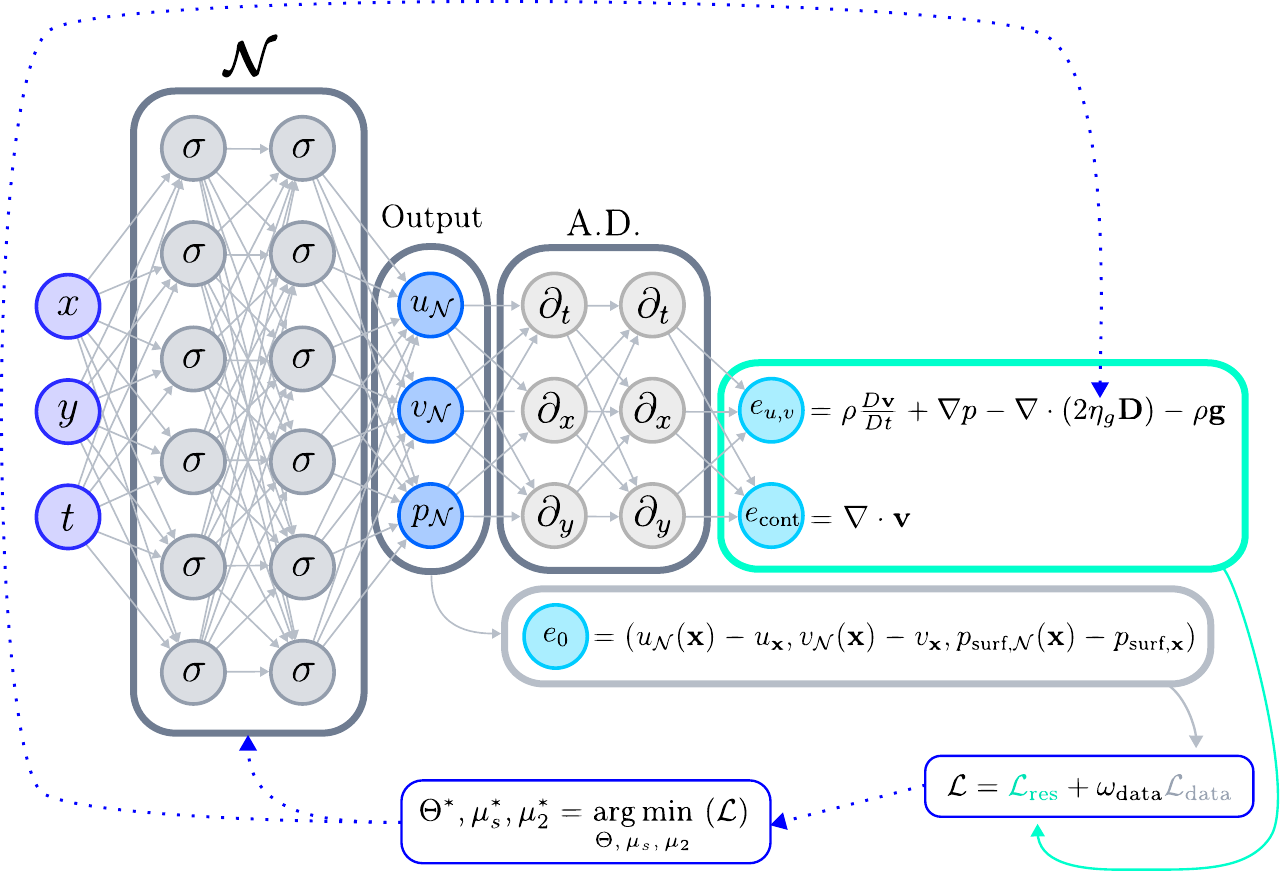}
    \caption{Physic informed neural network structure : a fully connected neural network take as input $\textrm{\textbf{x}}=(x,y,t) \in \mathbb{R}^{3}$ and predicts $\boldsymbol{\mathcal{N}}(\textrm{\textbf{x}}) = (u_{\mathcal{N}}(\textrm{\textbf{x}}),v_{\mathcal{N}}(\textrm{\textbf{x}}),p_{\mathcal{N}}(\textrm{\textbf{x}}))$. The residuals of the governing equations $e_{u,v}$ and $e_{\text{cont}}$ are computed by automatic differentiation and $e_0$ denotes the mismatch between the observational data ($u_\bfx$, $v_\bfx$, $p_{\text{surf},\bfx}$)
    %$(u_\text{obs},v_\text{obs})$
    and the ANN predictions, which are combined in the loss function $\mathcal{L}$. The trainable parameters ($\mu_s$, $\mu_2$) from the $\mu(I)$ rheology are initialized with the model, and their values are adjusted during training, similarly to the internal parameters $\Theta$ of the model.}
    \label{fig:PINN}
\end{figure*}

The expression of the optimization problem written as \eqref{eq:optivit} is correct for a flow reconstruction problem. However for a parameter identification problem, the problem has to be reformulated since e.g., $\mu_s$ and $\mu_2$ are two additional unknowns. In this case, the optimization problem reads: 
\begin{equation}
\label{eq:optivitid}
\bfTheta^{*},\mu^{*}_s,\mu_2^{*}= \underset{\bfTheta,\,\mu_s,\,\mu_2}{\argmin}  ~(\mathcal{L}_\text{res}+%\omega_\text{data} 
\mathcal{L}_\text{velo}),
\end{equation}
which is solved practically with the library Tensorflow by declaring the unknown parameters as trainable variables, the corresponding architecture is depicted in figure \ref{fig:PINN}.

\subsection{Training}

Several neural networks were trained, each corresponding to a flow regime considered. Each network consisted of 3 hidden layers with 250 neurons per layer, an architecture selected as it provided the lowest error percentages while keeping the training time reasonable. The \emph{swish} \cite{ramachandran2017searching} activation function was used for all layers except the last one where no activation function was selected.
At the beginning of the training process, the parameters $\mu_s$ and $\mu_2$ were initialized randomly for each training run. To ensure that their values remained physically meaningful throughout the optimization, both parameters were constrained within the range $[0,1]$ by applying a sigmoid transformation. 
For the optimisation process, we chose \emph{Adam} \cite{kingma2014adam} optimizer, a variant of the stochastic gradient descent algorithm. The number of epochs was fixed to 20000 epochs and the learning rate was decreasing from $5\times 10^{-4}$ to $1\times 10^{-5}$ during the total number of epochs. The training was done on GPU Nvidia RTX A5000, with a batch size of 512.

In the following sections, we quantitatively assess the accuracy of both the predicted fields and the identified parameters by computing the relative $L_2$ norm between the PINN predictions and the reference data on the test grid defined in section \ref{sec:datagen}. For the pressure field $p$, the relative $L_2$ error is defined as:
\begin{equation}\label{eq:errorL2}
    \epsilon_p =100 \times \sqrt{\dfrac{\displaystyle \sum_{\bfx \in \bfX} \norm{p_{\mathcal{N}, \text{test}}(\bfx_\text{test})-p_{\bfx_\text{test}}}^2}{\displaystyle \sum_{\bfx \in \bfX} \norm{p_{\bfx_\text{test}}}^2}},
\end{equation}
where $p_{\mathcal{N},\text{test}}(\bfx_\text{test})$ denotes the model prediction at point $\bfx_\text{test}$, and $p_{\bfx_\text{test}}$ is the corresponding reference value. The error is reported as a percentage for easier interpretation.
The same metric is also applied to evaluate the accuracy of the identified scalar parameters.

\section{Numerical results}
\label{sec:results}

In this section, we present the numerical results obtained with the PINN approach applied to the granular collapse configuration. While the main objective of this work is the identification of the $\mu(I)$-rheology parameters, we also show that the method enables the reconstruction of the pressure field from the velocity components $(u,v)$. To our knowledge there is no simple numerical procedure for this purpose such as the one proposed hereafter. Beyond its intrinsic interest, this reconstructed pressure provides an additional means of assessing the physical consistency and quality of the PINN solution. The results are therefore organized into three parts: parameter identification, pressure reconstruction as a consistency check of the learned solution, and an analysis of the robustness of the method with respect to measurement noise.

\subsection{Identification of $\mu_s$ and $\mu_2$}\label{sec:idmu2s}

In this section, we present the results obtained from training the PINN models, with the objective of simultaneously identifying the parameters $\mu_s$ and $\mu_2=\Delta\mu + \mu_s$.

Preliminary identification tests were performed using only the velocity dataset $\bfX^{\text{velo}}$, defined in section \ref{sec:datagen}, during training, which yielded unsatisfactory results in term of accuracy. 
Specifically, the relative error reached 53\% for the identification of $\mu_s$ and up to 100\% for $\mu_2$, indicating that the velocity field alone does not provide sufficient information to reliably constrain the rheological parameters. This led us to introduce and use a second dataset during training, the subset of points located at the interface between the granular media and the surrounding air, i.e. where $c\simeq 0.5$, and denoted as $\bfX^{\text{surf}}$. Pressure values are assigned exclusively to these interface points during training, as such quantities would typically be accessible in an experimental setting.
To incorporate this additional source of information, a new term was introduced in the loss function to account for the pressure values imposed at the free surface:
\begin{equation}
 \mathcal{L}_{p_{\text{surf}}} = \frac{1}{2\left|\bfX^{\text{surf}}\right|} \sum_{\bfx\in\bfX^{\text{surf}}}	\left|p_{ \text{surf}, \mathcal{N}}(\bfx) - p_{\text{surf},\bfx}\right|^{2} .
\end{equation}
This modification allows the network to explicitly incorporate pressure information at the granular–air interface, thereby improving the identifiability of the rheological parameters. Consequently, the overall optimization problem becomes:
\begin{equation} \label{eq:optprobid}
\bfTheta^{*},\mu^{*}_s,\mu_2^{*}= \underset{\bfTheta,\,\mu_s,\,\mu_2}{\argmin}  ~(\mathcal{L}_\text{res}+
%\omega_\text{data} 
\mathcal{L}_\text{data}),
\end{equation}
where $\mathcal{L}_\text{data} = \mathcal{L}_\text{velo} + \mathcal{L}_{p_{\text{surf}}}$. The inclusion of $\mathcal{L}_{p_{\text{surf}}}$ provides a physically meaningful constraint at the free surface, effectively coupling the velocity and pressure fields and leading to a substantial improvement in the accuracy of the identified parameters. In the following, we present the results obtained with the inclusion of interface pressure data.

\begin{figure}[h!]
    \centering
    \includegraphics[width=0.48\textwidth]{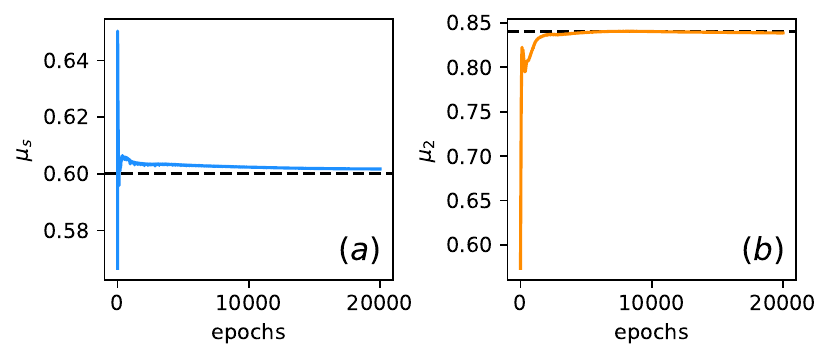}
    \vskip -0.3cm
    \caption{Evolution of the predicted value of (a) $\mu_s$ and (b) $\mu_2$ for the case of $\mu_s = 0.6$ and $\mu_2=0.84$. }
    \label{fig:evo_params}
\end{figure}

Figure \ref{fig:evo_params} shows the evolution of $\mu_s$ and $\mu_2$ during training for one representative configuration among the six studied ($\mu_s=0.6$ and $\mu_2=0.84$). The other flow configurations exhibit the same behavior and are therefore not shown here.
Both parameters rapidly converge toward stable values within the first few thousand epochs, with $\mu_s$ exhibiting the earliest and smoothest convergence, while $\mu_2$ shows only minor fluctuations before stabilizing. This systematic early convergence of $\mu_s$ is consistent with the distribution of inertial numbers in the granular collapse, which has a mean value on the order of $10^{-2}$. In this quasi-static regime, the $\mu(I)$ rheology varies most strongly with $\mu_s$, which corresponds to the value of $\mu$ as $I\rightarrow0$ and thus dominates the frictional response of the material (Fig. \ref{fig:muI}).

Table \ref{tab:id} summarizes the identification results of both rheological parameters  for the six different flow configurations considered. For each case, the corresponding identified values and their relative errors, denoted $\epsilon_{\mu_s}$ and $\epsilon_{\mu_2}$, are reported. The last column reports the relative errors $\epsilon_p$ for the pressure field reconstruction and will be discussed in the next section.

\begin{table}[h]
\centering
\begin{tabular}{|c|c|c|c|c|c|c|c|}
\hline
\multicolumn{3}{|c|}{Simulation} & \multicolumn{2}{|c|}{Identified} & \multicolumn{3}{|c|}{Error} \\
\multicolumn{3}{|c|}{parameters} & \multicolumn{2}{|c|}{parameters} & \multicolumn{3}{|c|}{(\%)} \\
\hline
$\mu_s$ & $\Delta \mu$ & $\mu_2$ &  $\mu_s$ &  $\mu_2$ &$\epsilon_{\mu_s}$ &$\epsilon_{\mu_2}$ &$\epsilon_{p}$  \\
\hline
0.2 & 0.12 & 0.32 &  0.201  &  0.318  &  0.59  &  0.61  &  0.88  \\      
0.2 & 0.24 & 0.44 &  0.201  &  0.438  &  0.57  &  0.41  &  0.98  \\                  
0.4 & 0.12 & 0.52 &  0.399  &  0.522  &  0.12  &  0.33  &  1.78  \\ 
0.4 & 0.24 & 0.64 &  0.403  &  0.622  &  0.80  &  2.87  &  1.59  \\ 
0.6 & 0.12 & 0.72 &  0.599  &  0.726  &  0.18  &  0.77  &  6.94  \\ 
0.6 & 0.24 & 0.84 &  0.602  &  0.839  &  0.28  &  0.15  &  6.43  \\ 
\hline   
\end{tabular}
\caption{Identification results of the rheological parameters $\mu_s$ and $\mu_2$, and reconstruction error of the pressure field.  The reference and identified values of $\mu_s$ and $\mu_2$ are reported together with their relative errors ($\epsilon_{\mu_s}$, $\epsilon_{\mu_2}$), and the relative pressure reconstruction error $\epsilon_p$.}
\label{tab:id} 
\end{table}

The identified values of $\mu_s$ and $\mu_2$ show an excellent agreement with the corresponding reference parameters across all test cases. The relative errors $\epsilon_{\mu_s}$ remain below 1\% for all configurations, indicating that the network accurately captures $\mu_s$. Similarly, the identified $\mu_2$ values are very close to the reference ones, with relative errors $\epsilon_{\mu_2}$ typically below 1\%, except for one case where the error reaches 2.87\%. This slight deviation at higher $\mu_2$ values may result from increased nonlinearity in the stress–strain relationship, which slightly challenges the identification process. Overall, the results confirm the reliability and robustness of the PINN framework for accurately identifying both rheological parameters.

\subsection{Reconstruction of the pressure field}
\label{sec:precon}

While measuring the velocity field can be experimentally achieved by mean of particle tracking velocimetry (PTV) techniques in the quasi-2D configuration \cite{gans2023collapse}, there is no experimental technique for measuring the pressure field \textit{in-situ}, which should be reconstructed. The PINN approach allows one to reconstruct the pressure by solving the optimization problem defined by \eqref{eq:optprobid}.  

The last column of table 1 reports the relative error $\epsilon_p$ associated with the reconstructed pressure field for each different flow considered. 
The pressure reconstruction error $\epsilon_p$ remains below 2\% for most configurations and increases moderately for the flows with the largest friction coefficients, where the pressure gradients become steeper.
Despite this, the overall accuracy remains satisfactory, indicating that the proposed PINN framework is capable of inferring the pressure field consistently with the identified rheological parameters.

\begin{figure*}[h!]
    \centering
    \includegraphics[scale=0.94]{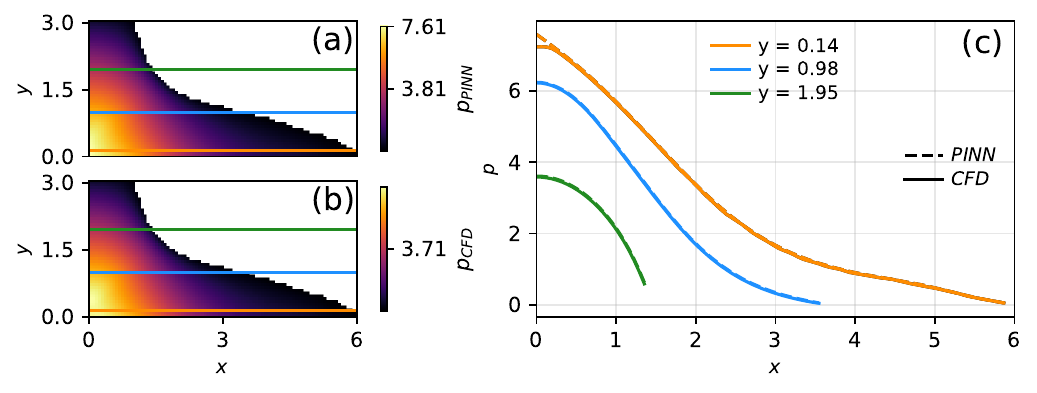}
    \vskip -0.4cm
    \caption{Comparison between the pressure field reconstructed by the PINN and the reference CFD solution for the case of $\mu_s=0.2$ and $\mu_2=0.32$ at $t=3.63$: (a) pressure field obtained from the PINN; (b) reference pressure field computed from CFD; (c) comparison between pressure profiles extracted at three transverse positions ($y=0.34$, $y=1.18$ and
    $y=2.15$).}
    \label{fig:champ_prof}
\end{figure*}

Figure \ref{fig:champ_prof} illustrates the pressure field reconstruction obtained with the PINN framework for the case of $\mu_s=0.2$ and $\mu_2=0.32$. Figure \ref{fig:champ_prof}a displays the pressure field predicted by the PINN for the quasi-2D configuration, while figure \ref{fig:champ_prof}b shows the corresponding reference CFD solution. 
The agreement between the two fields remains very good, with low discrepancies over most of the domain. This indicates that the reconstructed field accurately captures both the global pressure distribution and the spatial gradients.
Figure \ref{fig:champ_prof}c compares the pressure profiles at three transverse positions ($y=0.14$, $y=0.98$ and $y=1.95$), also indicated in figures \ref{fig:champ_prof}a and \ref{fig:champ_prof}b. The predicted profiles closely match the CFD data.
Minor deviations appear only in the non-flowing region where the velocity gradients vanish.

\begin{figure*}[h!]
    \centering
    \includegraphics[scale=0.86]{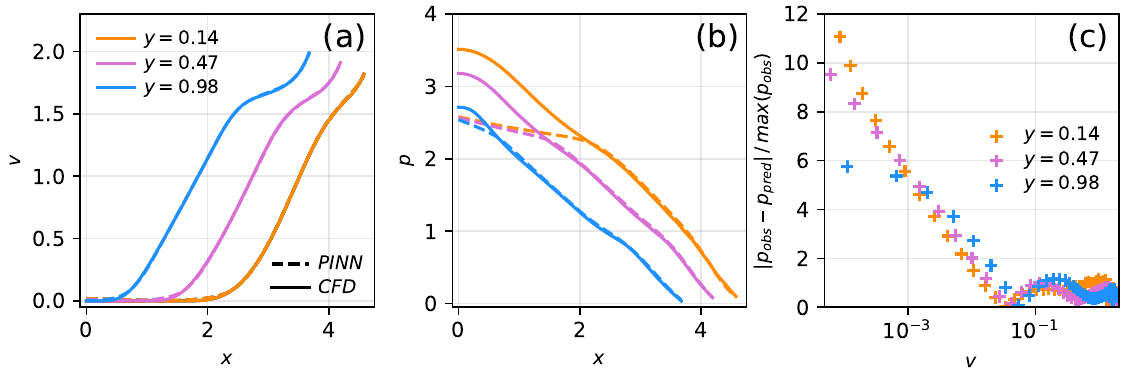}
    \vskip -0.3cm
    \caption{Comparison between the PINN predictions and the reference CFD solution for the case of $\mu_s=0.6$ and $\mu_2=0.84$ at $t=4.13$: (a) velocity profiles for three transverse positions ($y=0.14$, $y=0.47$ and $y=0.98$); (b) pressure profiles for the same three transverse positions; (c) relative difference between reconstructed and reference pressure profiles normalized by the maximum pressure, for the same three transverse positions.}
    \label{fig:prof_err}
\end{figure*}

For higher values of $\mu_s$, the behavior of the reconstruction remains consistent with these observations, as shown in figure \ref{fig:prof_err} for the case of $\mu_s=0.6$ and $\mu_2=0.84$. 
Figure \ref{fig:prof_err}a compares the velocity profiles obtained from the PINN and CFD for three transverse positions ($y=0.14$, $y=0.47$ and $y=0.98$), while figure \ref{fig:prof_err}b presents the corresponding pressure profiles. The agreement remains excellent in the flowing regions, confirming that the PINN accurately captures the coupling between velocity and pressure. The relative difference between reconstructed and reference pressure profiles, normalized by the maximum pressure (Fig. \ref{fig:prof_err}c), remains small across most of the domain but increases sharply where the velocity is constant and approaches zero. 
This increase in error is consistent with the known limitations of PINNs in regions where the fields gradients vanish. When the gradients become very small, the physical residuals involved in the loss function also approach zero, which reduces the sensitivity of the optimization process.  As a result, the network receives almost no corrective signal in these zones, leading to inaccurate reconstructions. Such behavior has already been reported in previous studies \cite{delcey2022physics}.  In the present case, the higher relative error observed near quasi-stagnant regions is therefore directly linked to the near-zero velocity gradients, which make these zones intrinsically difficult for PINNs to resolve.

Overall, these results confirm the robustness of the PINN framework for pressure reconstruction, while highlighting that the higher errors reported in table \ref{tab:id} originate from the expansion of non-flowing zones where the PINN has difficulty to accurately infer weak velocity fields due to their low magnitude and limited physical constraints in these regions.

\subsection{Identification with noise}

To evaluate the robustness of the proposed PINN-based identification approach, a sensitivity analysis was carried out with respect to measurement noise. In practical applications, velocity fields obtained from experiments are inevitably affected by uncertainties, which may influence the accuracy of the identified rheological parameters and the reconstructed pressure field. Assessing the performance of the method under such conditions is therefore essential to ensure its reliability when applied to real granular flow data.

\begin{table}[h]
\centering
\begin{tabular}{|c|c|c|c|c|c|c|c|}
\hline
\multirow{3}{*}{Noise (\%)} & \multicolumn{2}{c|}{Identified} & \multicolumn{3}{c|}{Error } \\
& \multicolumn{2}{c|}{parameters} & \multicolumn{3}{c|}{(\%)} \\ \cline{2-6}
& $\mu_s$ & $\mu_2$ & $\epsilon_{\mu_s}$ & $\epsilon_{\mu_2}$ & $\epsilon_{p}$ \\ 
\hline
0  &  0.604  & 0.833 & 0.67  & 0.78 & 7.11  \\      
20 &  0.606  & 0.830 & 1.01  & 1.24 & 9.71  \\      
80 &  0.586  & 0.953 & 2.34  & 13.4 & 14.5  \\
\hline
\end{tabular}
\caption{Influence of noise on the identification of the rheological parameters $\mu_s$ and $\mu_2$, and on the pressure reconstruction error $\epsilon_p$, for the reference case $\mu_s = 0.6$ and $\mu_2 = 0.84$. The added noise is Gaussian, with a mean value equal to the percentage of the maximum velocity magnitude indicated in the first column.}
\label{tab:bruit} 
\end{table}
Table \ref{tab:bruit} presents the influence of noise on the identification of the rheological parameters $\mu_s$ and $\mu_2$, as well as on the reconstructed pressure field, for the reference case $\mu_s = 0.6$ and $\mu_2 = 0.84$.
Noise was added to the velocity fields and to the pressure at the interface. In both cases, the noise followed a Gaussian distribution. When applied to the velocity field, its mean value was set as a given percentage of the maximum absolute velocity magnitude; similarly, when applied to the interface pressure, the mean value corresponded to a percentage of the maximum absolute interface pressure. The noise levels tested are indicated in the first column of the table.

As shown, the identification procedure remains stable and accurate for moderate noise levels. When up to 20\% noise is added, the relative errors $\epsilon_{\mu_s}$ and $\epsilon_{\mu_2}$ remain below 1.5\%, indicating that the network maintains a good estimation capability despite perturbations in the data. However, when the noise level reaches 80\%, the accuracy degrades more noticeably: $\epsilon_{\mu_s}$ increases to 2.34\% and $\epsilon_{\mu_2}$ to 13.4\%. The corresponding pressure reconstruction error $\epsilon_p$ also rises significantly, from 7.1\% in the noise-free case to 14.5\%.
These results demonstrate the robustness of the proposed PINN approach under noise conditions, while also highlighting a sensitivity of the $\mu_2$ identification and pressure reconstruction to high noise levels.

%\clearpage

\section{Conclusion}
\label{sec:conclu}

In this work, we have applied, for the first time, the Physics-Informed Neural Network (PINN) framework to identify the parameters of the $\mu(I)$-rheology for granular materials. 
This approach introduces a new paradigm for rheometry, circumventing the need for specialized and often costly experimental setups by relying solely on flow kinematics. Specifically, we have demonstrated that PINNs trained on synthetic data from a granular column collapse can accurately recover the friction coefficients $\mu_s$ and $\mu_2$ from the velocity field alone.

Under noise-free conditions, the identification results confirm the strong potential of this approach. When $\mu_s$ and $\mu_2$ are identified, the recovered values are highly accurate, with relative errors below a few percent. 
The robustness of the method was also evaluated under increasing levels of synthetic noise. The PINN retained good predictive capabilities under moderate perturbations of the velocity field and was still able to identify the key rheological parameters with reasonable accuracy even when subjected to significant Gaussian noise. Only a moderate sensitivity of $\mu_2$ and of the reconstructed pressure field was observed at high noise levels. This robustness strongly supports the applicability of the proposed framework to experimental velocity data, which are inherently noisy.

Building on these encouraging results, the next natural step is to apply the proposed framework to experimental measurements in order to assess its performance under realistic conditions and move toward data-driven rheometry directly informed by physical experiments. Although this study is purely numerical, the granular column collapse configuration has well-documented experimental counterparts in quasi-2D geometries, where velocity fields can be measured using Particle Tracking Velocimetry (PTV) or Particle Image Velocimetry (PIV) \cite{gans2023collapse}. Consequently, the PINN methodology appears directly compatible with existing experimental techniques and could enable quantitative rheological characterization of granular materials from experimentally accessible kinematic data.

In the present study, the parameters $\mu_s$ and $\mu_2$ were identified while the inertial parameter $I_0$ was kept fixed. This strategy proved effective and led to highly accurate reconstructions. However, in principle, identifying $I_0$ alongside $\mu_s$ and $\mu_2$ would provide a more complete characterization of the $\mu(I)$ rheology. A preliminary attempt was made to jointly identify all three parameters using a straightforward optimization approach, but it did not yield satisfactory results. We hypothesize that this limitation arises because $I_0$ induces only minor variations in the rheological law $\mu(I)$ over the range of inertial numbers sampled by the granular collapse configuration. 
Considering all six datasets explored in this study, the inertial numbers are generally very small for the majority of the material, typically on the order of $10^{-2}$, with median values around 0.02–0.03 and 95\% of the values are between 0 and 0.09. Maximum values occasionally reach a few units, while minimum values are essentially zero. This indicates that most of the flow occurs in a low-inertia regime where $\mu(I)$ is relatively insensitive to $I_0$, which explains the difficulty in accurately identifying this parameter. Future work will focus on exploring alternative flow configurations that exhibit a broader and more uniformly distributed range of inertial numbers, which may enable a more reliable identification of $I_0$.

Overall, the results presented in this work highlight the feasibility and potential of using PINNs for data-driven identification of rheological parameters in granular flows. Beyond the numerical demonstration, the proposed PINN-based framework has the potential to transform how rheological parameters are determined in industrial and research environments. Traditionally, estimating the frictional properties of granular media requires complex rheometric devices, highly controlled conditions, or costly large-scale experiments. In contrast, the approach developed here extracts the material parameters directly from flow kinematics, relying only on velocity fields and interface pressure data that are experimentally accessible in simple quasi-2D column collapse configurations. This capability opens new perspectives for a hybrid form of rheometry that combines physical modeling, data assimilation, and artificial intelligence to infer granular material properties from flow observations under realistic conditions.

\section*{Acknowledgements}

A part of this study is conducted in the framework of the Agence Nationale de la Recherche (Project
‘Pinn’terfaces’ ANR-23-CE23-0020).

\section*{Author contributions (CRediT)}

\textbf{Barbara Baldoni} : Conceptualization, Data Curation, Investigation, Methodology, Software, Validation, Visualization, Writing - Original Draft.
\textbf{Mickaël Delcey} : Methodology, Software.
\textbf{Yoann Cheny} : Conceptualization, Funding acquisition, Methodology, Supervision, Writing - Original Draft.
\textbf{Adrien Gans} : Data Curation, Methodology, Writing – Review \& Editing.
\textbf{Mathieu Jenny} : Writing – Review \& Editing.
\textbf{Sébastien Kiesgen de Richter} : Writing – Review \& Editing.

\section*{Declaration of competing interests}
The authors have nothing to declare.

\section*{Declaration of generative AI in the manuscript preparation process}
During the preparation of this work the author(s) used ChatGPT in order to enhance the clarity and quality of the text. After using this tool/service, the author(s) reviewed and edited the content as needed and take(s) full responsibility for the content of the published article.

\bibliographystyle{elsarticle-num} 
\bibliography{bib}
\end{document}